\documentclass[a4paper,fleqn]{cas-dc}

\usepackage[numbers]{natbib}
\usepackage{graphicx}
\usepackage{dcolumn}
\usepackage{bm}
\usepackage{multirow}
\usepackage{csquotes}
\usepackage{tabu} 
\usepackage{graphicx}
\usepackage{dcolumn}
\usepackage{bm}
\usepackage{caption}
\usepackage{booktabs}
\usepackage{float}
\restylefloat{table}

\begin{document}
\let\WriteBookmarks\relax

\title [mode = title]{Vertical electrolyte transistor operating at very low voltage and high current density}

\author[1]{Keli Fabiana Seidel}

\cormark[1]
\fnmark[1]

\address[1]{Universidade Tecnol\'ogica Federal do Paran\'a -- UTFPR, Av. Sete de Setembro, 3165 - 80230-901 - Curitiba - Brazil}
\address[1]{Humboldt-Universit\"at zu Berlin, Institut f\"ur Physik, Brook-Taylor-Str. 6, 12489 - Berlin - Germany }

\cortext[cor1]{corresponding author: \\E-mail address: keliseidel@utfpr.edu.br}


\begin{abstract}
In this work it is reported a vertical electrolyte transistor (VET) whose structure is based on stacked layers as described below: bottom contact (source or drain) $\rightarrow$ channel $\rightarrow$ permeable intermediate electrode (drain or source) $\rightarrow$ ion gel (electrolyte gate dielectric) $\rightarrow$ gate top contact. This VET depicts versatility to work as Electrolyte-Gated Vertical Organic Field Effect Transistor (Electrolyte-Gated VOFET) or Vertical Organic Electrochemical Transistor (VOECT) as never reported before. The distinction of these operation modes is regarding to the transistor transconductance that occurs due to induced charge carriers or ionic current, respectively. Both modes of operation show that this VET is able to work at very low voltage range and drive a high current density. These observed features make VETs a good candidate for applications in iontronic devices, bio-sensors and/or very low power optoelectronic circuits.
\end{abstract}




\maketitle
\section{Introduction}
\par Electrolyte-gated organic field effect transistors (EGOFETs) form a featured transistor class with great potential for application in low power (bio)-electronics \cite{EGOFET_DNA_detection, Emil_EGOFET_SPIE_1,EGOFET_biosensor_review, bioelectronics_Malliaras, Water_gated_organic_transistors_Biscarini}. In the EGOFET device architecture, the organic semiconductor is in direct contact with the analyte. The gate-induced charge occurs due to the electric double layer formed at the semiconductor/electrolyte interface that can be described as a Helmholtz layer \cite{water_gate_transistor_Horowitz_1,Emil_EGOFET_SPIE_1,Emil_EGOFET_APL_2,Lee_ion_gel_2012}. This electrolyte layer can be formed by an ionic liquid \cite{water_gate_transistor_Horowitz_1} or ion gel \cite{Lee_ion_gel_2012} and both provide higher gate capacitance (up to $\sim\,1000$ higher) compared to the non-electrolytic dielectrics \cite{EGOFET_biosensor_review}. That enables EGOFETs to operate at low voltages ($<0.5\,\rm{V}$) \cite{water_gate_transistor_Horowitz_1, Emil_EGOFET_SPIE_1}.
\par Another class of transistors able to operate at very low voltage range ($\sim 1.5\,\rm{V}$) is the class of vertical organic field effect transistors (VOFETs) \cite{Seidel2018,seidel2013,graeff_Ag_NW}. In these types of devices all the layers are stacked in order to form a diode cell on a capacitive cell (or vice versa) \cite{Seidel2018,yang_2004}. Its channel length is only a few nanometers, allowing an output current to be obtained at a low drain voltage for any gate voltage applied. Compared to EGOFETs, VOFETs show higher current density at equivalent vol\-tage ranges and fast response times by up to three orders of magnitude \cite{graeff_Ag_NW,EGOFET_frequency}.
\par In this work, we demonstrate a vertical electrolyte transistor (VET), composed of stacked layers as depicted in Figure \ref{VEGOFET_architecture}. This architecture is based on a blend architecture/struc\-ture of the two well-known devices introduced before: (i) solid state EGOFETs, with an ion gel as electrolyte gate dielectric and, (ii) VOFETs, in which all the layers are stacked. As far as the author know, there are just two similar VET already reported in the literature named as Electrolyte-Gated Vertical Organic Transistor (VOT) in 2018 \cite{VET_2018} and Electroly\-te-Gated VOFETs in 2019 \cite{VET_2_Lenz}. Here, the choose of this new name, Vertical Electrolyte Transistor (VET), is an attempt to associate it only with its architecture/structure. In the sequence, after the charge carriers transport has been analysed, similar name to the existing in literature \cite{VET_2_Lenz} will be attributed for transistors with transconductance generated by field effect only. Since there are more than one possibility to explore the transconductance in VETs and it has never been reported before in literature, that is one of the main goal in this paper to describe differenciated charge carriers regimes in VETs. All the new defined names have been choose in an attempt to keep in agreement with the architectures/structures largely studied before in literature \cite{yang_2004,water_gate_transistor_Horowitz_1,Organic_electrochemical_transistors_Malliaras,VET_2_Lenz}.
\par The results presented here are with respect to the versatility of this structure, which is able to be implemented in two different transistor principles, namely: (i) Electro\-lyte-Gated Vertical Organic Field Effect transistor (Electro\-lyte-Gated VOFET), where the modulation of the output current occurs due to the induced charge carriers in the channel or, (ii) Vertical Organic Electrochemical Transistor (VOECT), where the modulation of the output current occurs due to the ions that diffuse into the channel. Both transistor operations allow very low voltage operation ranges with high current density.
\section{Experimental section}
\par The VET structure consists of: (a) bottom contact -- indium tin oxide (ITO); (b) channel -- poly(3-hexylthiophene-2,5-diyl) regioregular (P3HT) from chlorobenzene solution ($\sim 150\, \rm{nm}$); (c) intermediate permeable electrode -- a blend of silver nanowires-(Ag-NWs) with Poly(methyl methacrylate) (PMMA) deposited by drop casting; (d) electrolyte gate dielectric - deposition of the ion gel produced from a process called \textquote{cut and stick} as described by K. H. Lee et al. \cite{Lee_ion_gel_2012}; (e) top gate contact -- \textquote{cut and stick} deposition of gold (Au) foil 
with $\sim 7\,\rm{nm}$. The transistor active area is $\sim 9\,\rm{mm^2}$.
\begin{figure}[]
\begin{center}
\includegraphics[width=0.97\columnwidth]{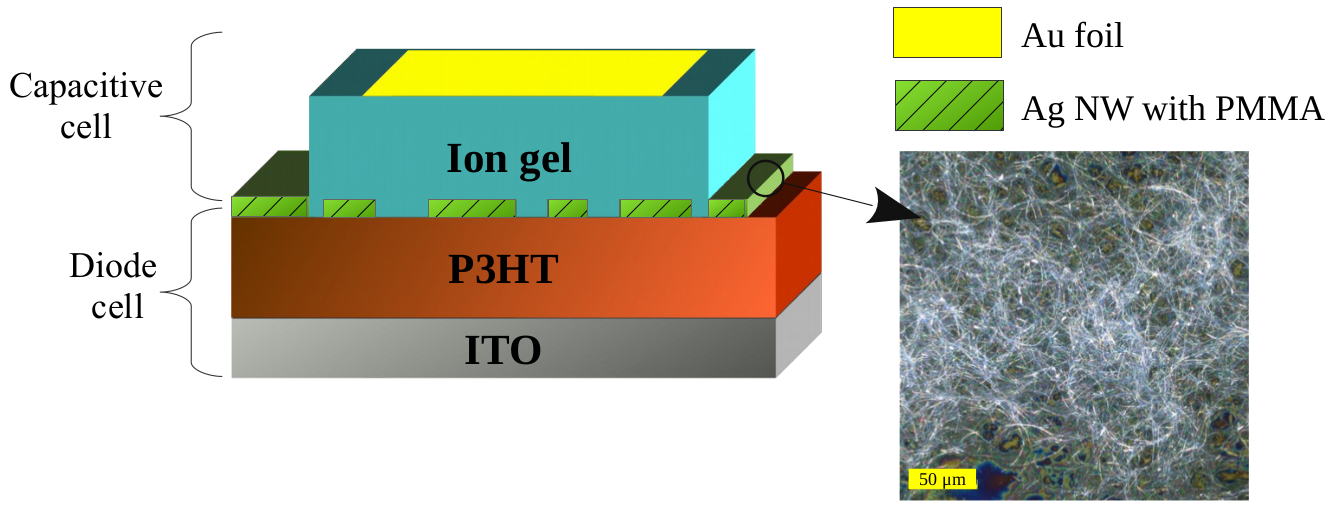}
\end{center}
\caption{\label{VEGOFET_architecture} VET architecture based on a capacitive cell on a diode cell. Device structure: ITO (source or drain), P3HT (channel), Ag-NW with PMMA (intermediate permeable electrode / drain or source), ion gel (electrolyte gate dielectric) and Au foil (gate contact). The microscope image depicts the Ag-NW with PMMA deposited on P3HT with good connection/distribution in the plane between the NWs.}
\end{figure}
\par ITO films ($100\,\rm{nm}$, $20\,\rm{{\Omega}/sq}$) on glass substrates were acquired from Ossila. The ITO substrates were cleaned in a sequence of acetone, isopropanol and water in an ultrasonic bath. Afterwards, the substrates were cleaned for $15\,\rm{minutes}$ in an UV ozone cleaner (Ossila). Immediately after this cleaning step, the P3HT film was deposited on the ITO electrodes. 
\par Regioregular poly(3-hexylthiophene-2,5-diyl) (P3HT) was supplied from Rieke Metals. It was deposited from chlorobenzene solution ($10\,\rm{mg/mL}$) by dynamic spin coating twice at $60\,\rm{\mu L}$ at $600\,\rm{rpm}$. The P3HT film thickness was $\sim 150\, \rm{nm}$. The film was annealed at $80^o\rm{C}$ for $30\,\rm{min}$ in inert gas atmosphere.
\par To produce the intermediate electrode (IE), a blend of Ag-NWs with PMMA was prepared to obtain the permeable IE. PMMA was added to the solution to increase the solution's surface tension avoiding the droplet spread on the P3HT film edges forming a short circuit with the bottom contact. The PMMA solution was prepared according to literature \cite{PMMA_ethanol_water} using $1\, \rm{mg/mL}$ concentration. The co-solvent used was $1.0\, \rm{mL}$ ethanol/water (80/20 wt-\%). This solution was stirred at $80\,\rm{^oC}$ overnight. $400\, \rm{\mu L}$  Ag-NW from isopropanol was then mixed with $200\, \rm{\mu L}$ PMMA solution and stirred at $80\,\rm{^oC}$ for $5\, \rm{min}$. $10\, \rm{\mu L}$ of this blend solution was then drop cast on the P3HT film. Figure \ref{VEGOFET_architecture} depicts a microscope image of Ag-NW/PMMA IE with good in-plane interconnection between the NWs forming a network.
\par The ion gel was prepared at the same concentration as described by K. H. Lee et al \cite{Lee_ion_gel_2012} where it was reported a capacitance of $10\, \rm{\mu}  F \,cm^{-2}$ for $10\, \rm{\mu} m$ thick ion gel sandwiched between two gold electrodes. Its solution was prepared from acetone with poly(vinylidene fluoride-co-hexafluoropropyl\-ene), P(VDF-HFP); and the ionic liquid 1-ethyl-3-methyl\-imidazolium  bis(trifluoromethylsulfonyl) ami\-de. It was stirred for $30\,\rm{min}$ at $55\rm{^oC}$. $100\,{\mu L}$ of the solution was then drop cast on a clean glass slide and the resulting film was stored in inert gas atmosphere. Following the same procedure described in literature \cite{Lee_ion_gel_2012}, the ion gel film was ``cut and stuck" on the Ag-NW intermediate electrode followed by a ``cut and stuck" gold foil to form the gate contact (see Figure \ref{VEGOFET_architecture}).
\section{Results and Discussion}
\par The electronic properties of the produced VET was analysed from the transfer and output characteristics curves. The transistor was tested for two different very low gate voltage ranges, as depicted in Table \ref{table}: (Range A) $-\,0.3\,V < V_{g,A} < +\,0.3\,V$ and (Range B) $-\,0.6\,V < V_{g,B} < +\,0.6\,V$, for ambipolar $V_{ds}$. According to the applied voltage bias and its magnitude, the transistor transconductance occurs due to the induced charge carriers or ionic current. The analysis was done for ambipolar $V_{ds}$ with applied voltage at the bottom electrode (ITO) with reference to the IE. Its structure presents a built-in voltage close to zero. The work function of ITO is $\sim 4.8\,\rm{eV}$ and Ag-NW $\sim 4.6\,\rm{eV}$ \cite{silver_nanowires_WF}. As P3HT HOMO is at $\sim 5\,\rm{eV}$, hole injection from ITO or Ag-NW into P3HT is enabled at low voltages \cite{zero_built_in_potential_1,zero_built_in_potential_2}.
\begin{table}[H]
    \centering
\begin{tabular}{|c|c|}
 \hline
 \textbf{Gate Voltage Range} &  \textbf{$\Delta V_g$}\\ 
 \hline\hline
 Range A ($V_{g,A}$) &  $-0.3\,\rm{V} < V_{g,A} < + 0.3\,\rm{V}$\\ 
 \hline
 Range B ($V_{g,B}$) &  $-0.6\,\rm{V} < V_{g,B} < + 0.6\,\rm{V}$\\
 \hline
\end{tabular}
\caption{Gate voltage ranges applied on VET.}
\label{table}
\end{table}
\par
\begin{figure}[]
\begin{center}
\includegraphics[width=7.9cm,height=4.4cm]{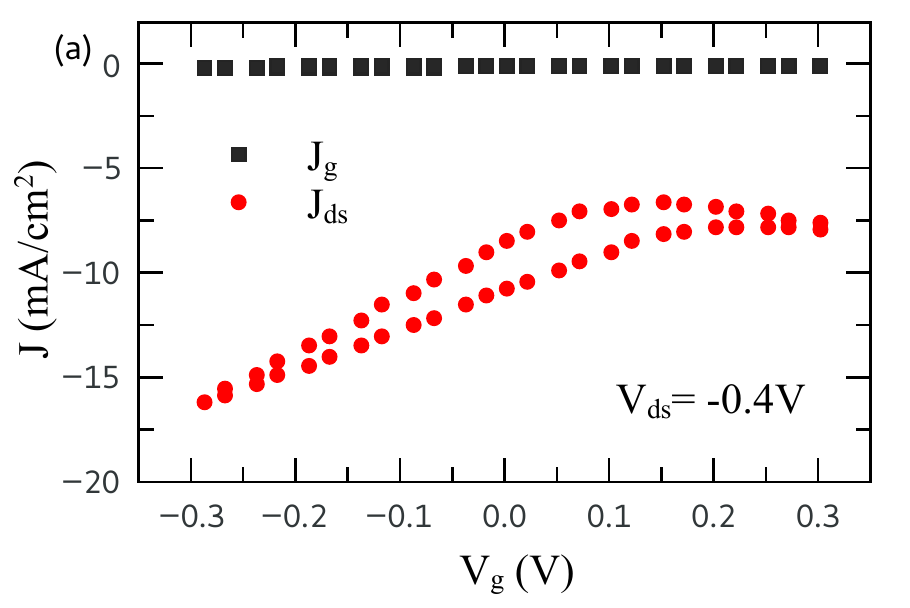}
\includegraphics[width=7.9cm,height=4.4cm]{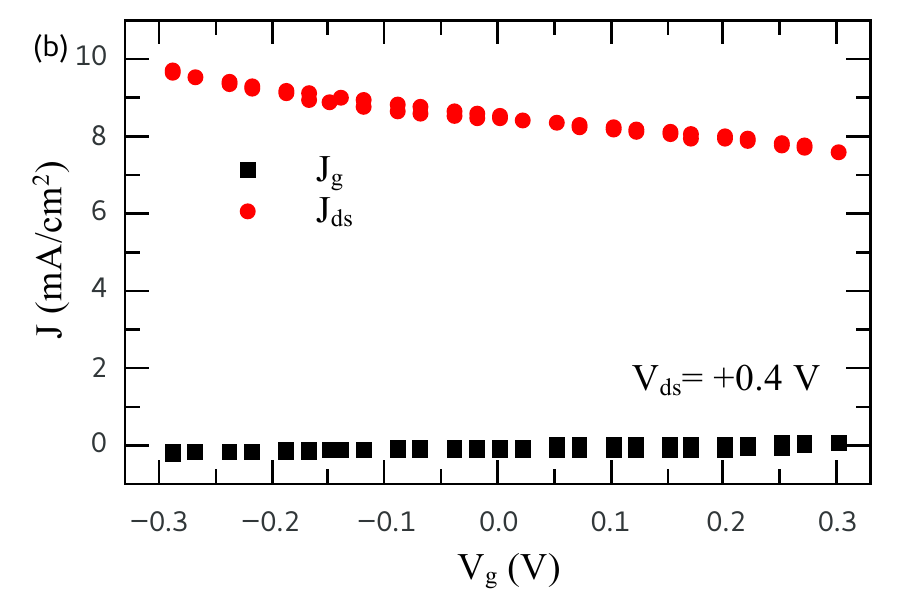}
\includegraphics[width=8.1cm,height=4.4cm]{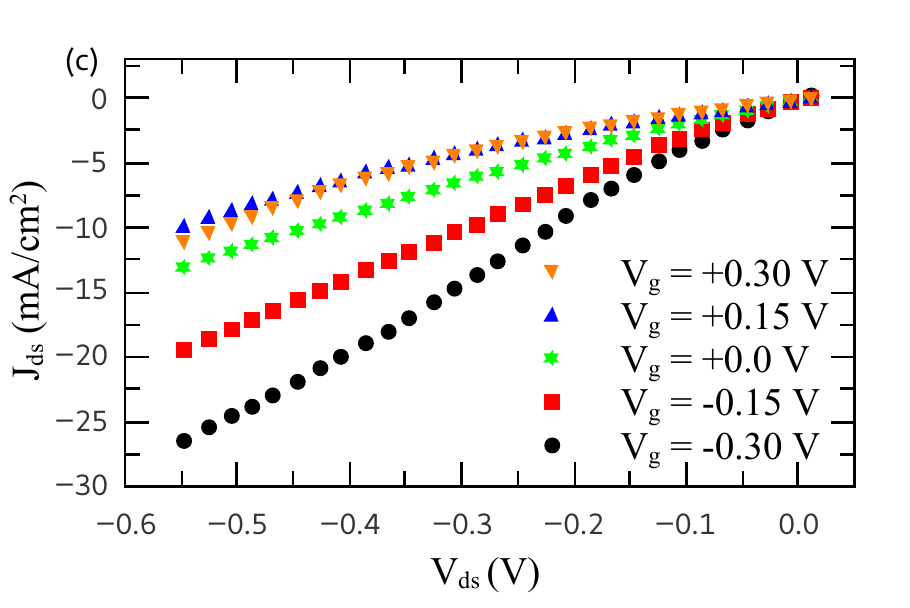}
\end{center}
\caption{\label{characterization_range_A}  VET electrical characterisation for gate voltage range A with unipolar output current -- (a) Transfer curve for ambipolar gate biasing de\-picting the output current density ($J_{ds}$--red circles) and leakage current density ($J_{g}$--black squares) for $V_{ds}=-0.4\,\rm{V}$. (b) Characteristic curve for unipolar current density modulated by ambipolar gate biasing.}
\end{figure}

 The transfer curve ($J_{ds}\times V_g$) for $V_{ds}=-0.4\,\rm{V}$ and ambipolar gate biasing (Range A) is depicted in Figu\-re \ref{characterization_range_A}(a). The transconductance ($g=\Delta I_{ds} / \Delta V_g$) for negative gate bias is more pronounced than for positive gate bias. The black-squares in Figure \ref{characterization_range_A}(a) depict the very low leakage current density ($J_g$, the current between gate and IE), which is at least three orders of magnitude lower than the output current density ($J_{ds}$ - red circles). 
\par Figure \ref{characterization_range_A}(b) shows the transfer curve for $V_{ds}=+0.4\,\rm{V}$ (Range A) with an almost negligible transconductance. In this electrical configuration, holes are injected from the bottom electrode (ITO) while the applied gate voltage changes the energy barrier in the anode-IE/channel interface. However, as the anode energy barrier is too high, this low voltage applied is not enough to improve the charge carrier injection and no significant modulation is observed.
\par
\begin{figure}[]
\begin{center}
\includegraphics[width=0.75\columnwidth]{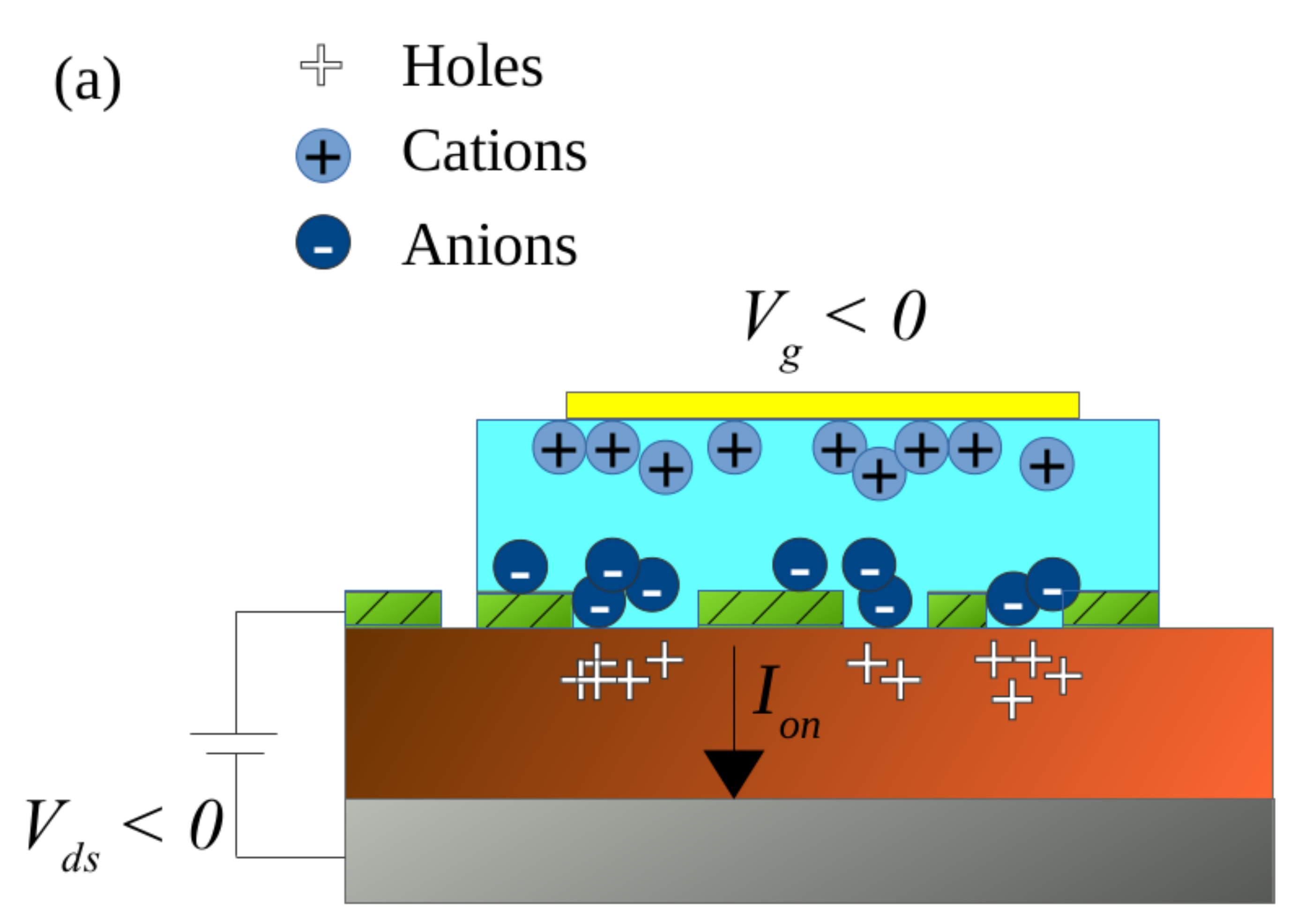}
\vspace{0.3cm}
\includegraphics[width=0.79\columnwidth]{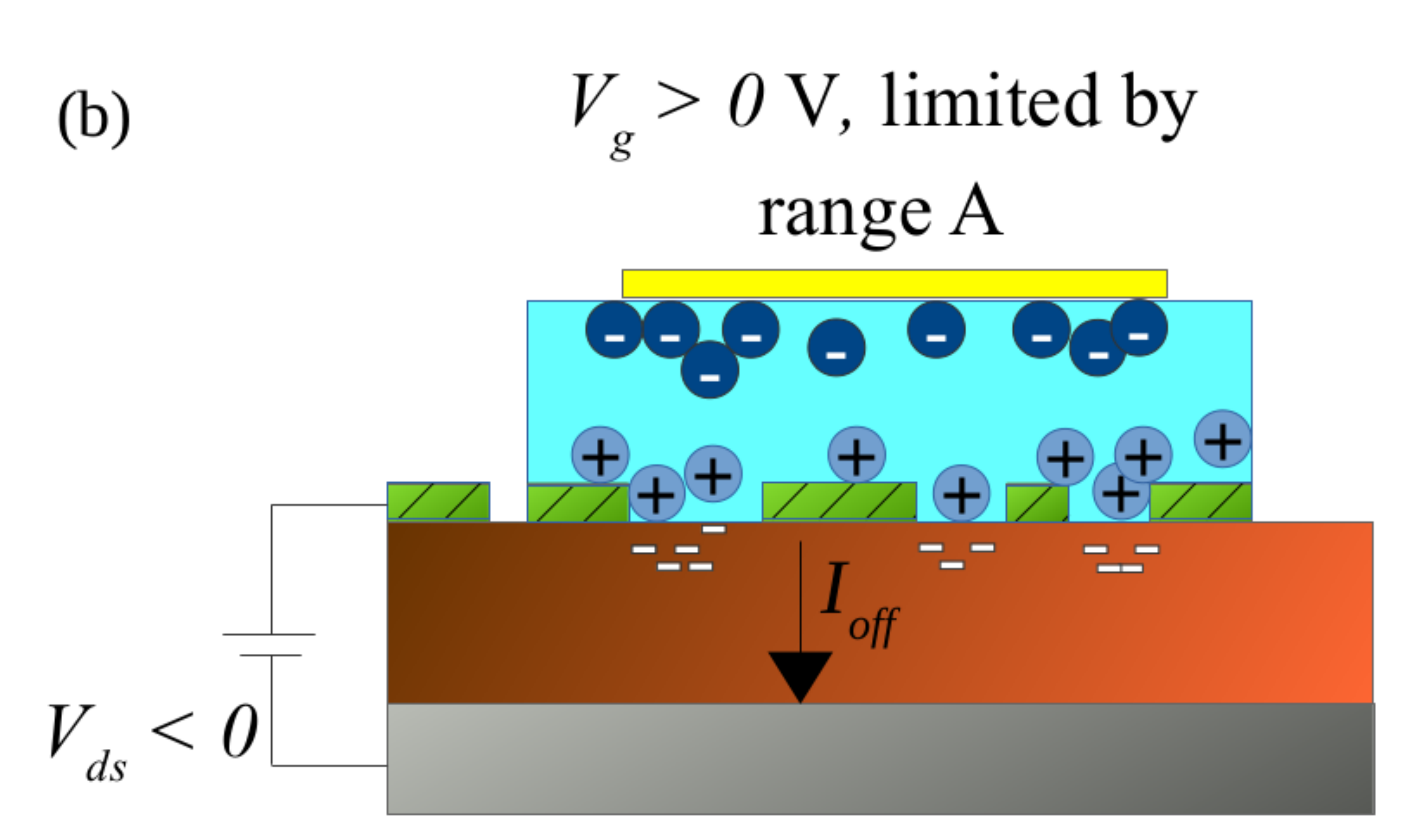}
\includegraphics[width=0.79\columnwidth]{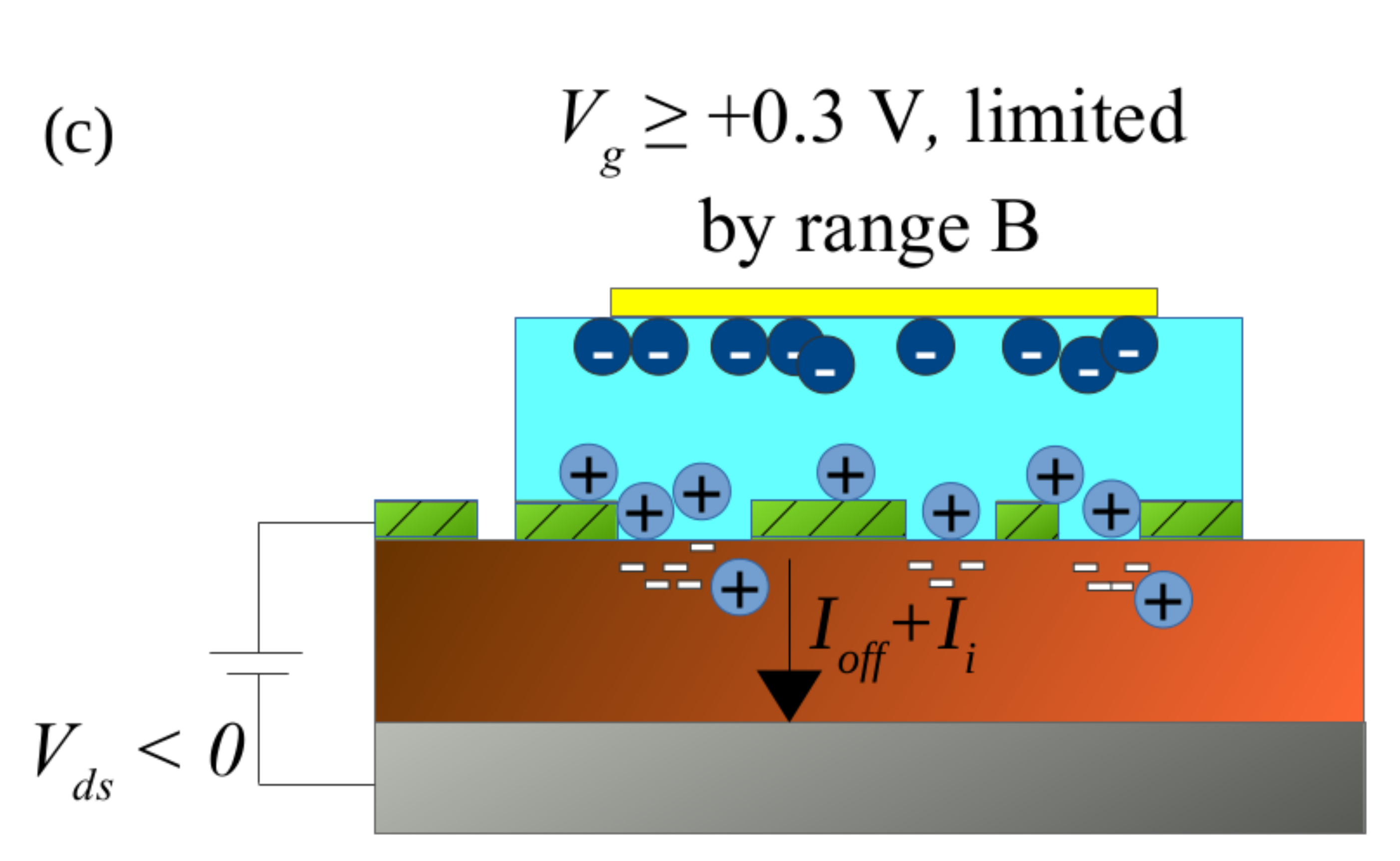}
\end{center}
\caption{\label{polarization_VEGOFET} Illustration of charge carrier distribution for this VET when: (a) $V_{ds}<0\,\rm{V}$ and $V_g<0\,\rm{V}$ forming the on-state ($I_{on}$= on-state current). The Helmholtz double layer is formed at the interface of the IE pores and the organic semiconductor. (b) $V_{ds}<0\,\rm{V}$ and $V_g>0\,\rm{V}$ (limited by the voltage range A) forming the off-state ($I_{\rm{off}}$= off-state current). (c) $V_{ds}<0\, \rm{V}$ and $V_g \geq  +0.3\,\rm{V}$ (limited by the voltage range B), where $I_i=$ ionic current.}
\end{figure}

 The characteristic curve ($J_{ds}\times V_{ds}$) for negative $V_{ds}$ and ambipolar gate biasing is depicted in Figure \ref{characterization_range_A}(c). For negative gate biasing setup, holes are induced in the channel interface through the pores of the IE as illustrated in Figu\-re \ref{polarization_VEGOFET}(a). This charge carrier polarization increases the output current intensity. Note, when gate voltage is applied, charge carriers are induced just at the interface channel/IE pores. The output current is therefore a sum of: (1) The current from the IE portion without pores that are the regions not affected by the applied gate voltage because the electric field is shielded by the Ag-NW. That is the same output current formed when $V_g=0\,\rm{V}$ and; (2) The portion related to the pores regions where the gate electric field permeates. The energy barrier changes take place at these regions producing the transistor modulation as already described in VOFET literature \cite{Seidel2018,tessler_theory}.
\par In this VET, the output current modula\-tion profile as well as its behavior are similar to that observed in the broad VOFET literature \cite{seidel2013,graeff_Ag_NW,Seidel2018}. The saturation regime in the output current is not reached since the channel length is too short and the low injection energy barrier. These same properties also result in transistors with a non-negligible output current, even for $V_g=0V$. However, an advantage of the architecture presented herein over VOFETs is the higher capacitance of the ion gel dielectric layer when compared to a dielectric layer from VOFETs. This allows modulation to be observed for an applied gate voltage of just up to $V_g=\pm |0.3|\,\rm{V}$, a range not previously shown in VOFETs. The ion gel high capacitance is due to the Helmholtz double layer (induced charges carriers/ions double layer) as shown in Figure \ref{polarization_VEGOFET}, a phenomenon widely studied in EGOFET literature \cite{Emil_EGOFET_SPIE_1,EGOFET_model}. 
\par Within this first analysed voltage range A, our VET pre\-sents transconductance due to the induced charge carriers in the channel/IE pores. Because of its operation mode, for the gate voltage range A we named this device as Electrolyte-Gated VOFET. Then, for negative $V_{ds}$ bias, it is possible to define the Electrolyte-Gated VOFET on-state with a negative gate voltage setup (as in Figure \ref{polarization_VEGOFET}(a)). However, when a positive gate voltage range is applied, 
electrons are induced at the interface channel/IE pores that will close the transport in the pores (see Figure \ref{polarization_VEGOFET}(b)). In these regions the current is decreased forming the Electrolyte-Gated VOFET off-state. As depicted in Figure \ref{characterization_range_A}(c), the output current decreases when $V_{g,A}=+0.15 V$. Then, the transport in the pores regions are closed but the output current keeps with the same density in the regions without pores where there are no induced charge carries due to the applied gate voltage and the interface energy barrier is not affected. This behavior has already been widely explained in VOFETs lite\-rature where this gate voltage bias, able to create the off-state, is named backward gate voltage \cite{Seidel2018, tessler_theory}.
\par This VET was tested also for $V_{g,A}>0.15\,\rm{V}$. However the output current increases again for $V_{g,A}=+0.30\,\rm{V}$. For this electrical setup an ionic current ($I_i$) appears through the channel/IE interface and this $I_i$ will generate the modulation making the output current density increases once again (see Figure \ref{polarization_VEGOFET}(c)). Therefore, this electrical configuration does not present the lowest output current intensity for this VET since this result is the sum of: (i) the current from the regions where there is no IE pores with intensity equivalent to the off-state current ($I_{\rm{off}}$) added to,  (ii) the ionic current ($I_{i}$) from the IE pores region. Due to the ions diffusion through the semiconductor, this behavior is similar to that observed in Organic Electrochemical Transistors (OECTs) \cite{Organic_electrochemical_transistors_Malliaras} and its transconductance does not depend just on the induced charge carriers. The discussion on charge carriers transport associated with the ionic current will be
retaked soon after. Based on transconductance due to only induced charge carriers, the On-Off ratio for the present Electrolyte-Gated VOFET (Range A) is defined to $V_{g,A}=-0.3V$ (On-state) and $V_{g,A}=+0.15V$ (Off-state). 
\par
\begin{figure}[]
    \centering
\includegraphics[width=7.9cm,height=4.4cm]{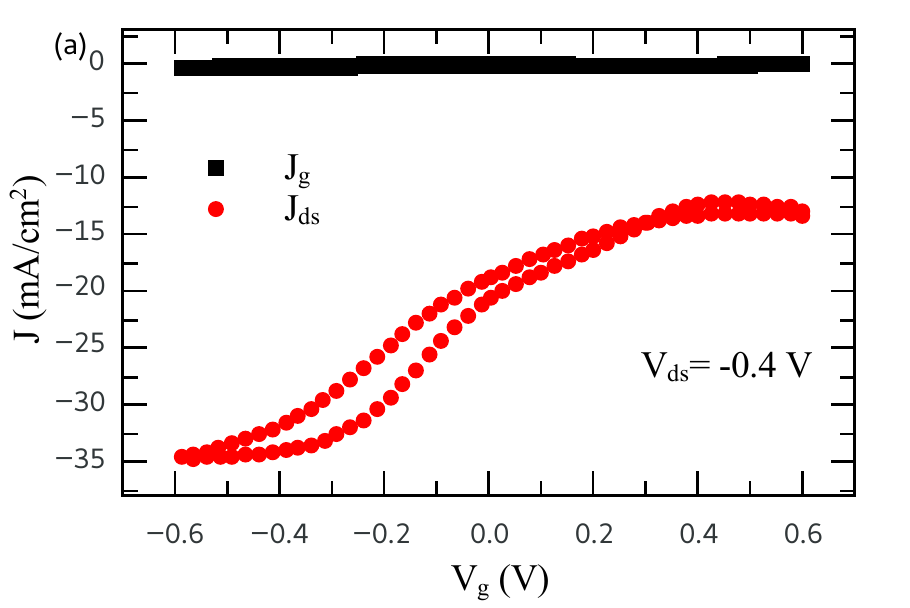}
\includegraphics[width=7.8cm,height=4.4cm]{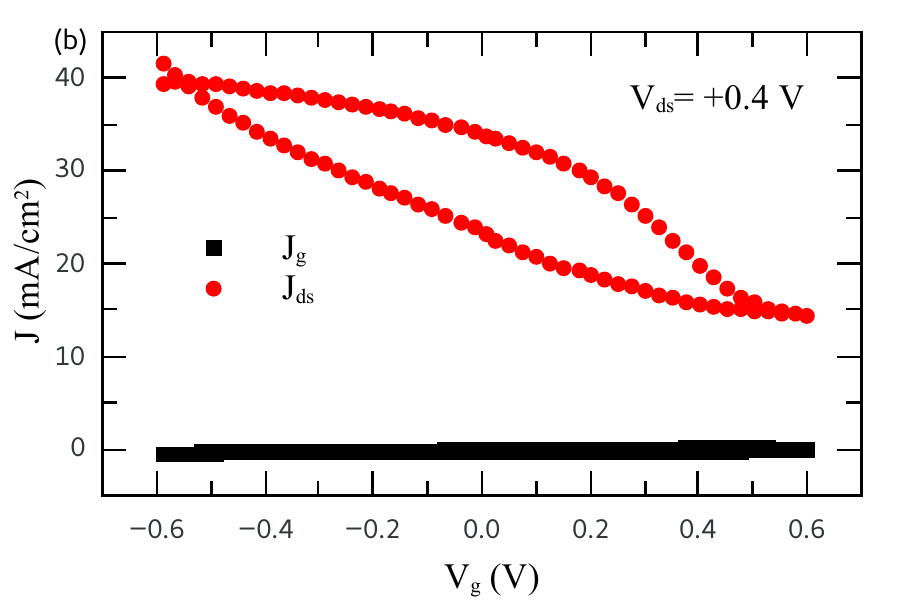}
\includegraphics[width=7.8cm,height=4.4cm]{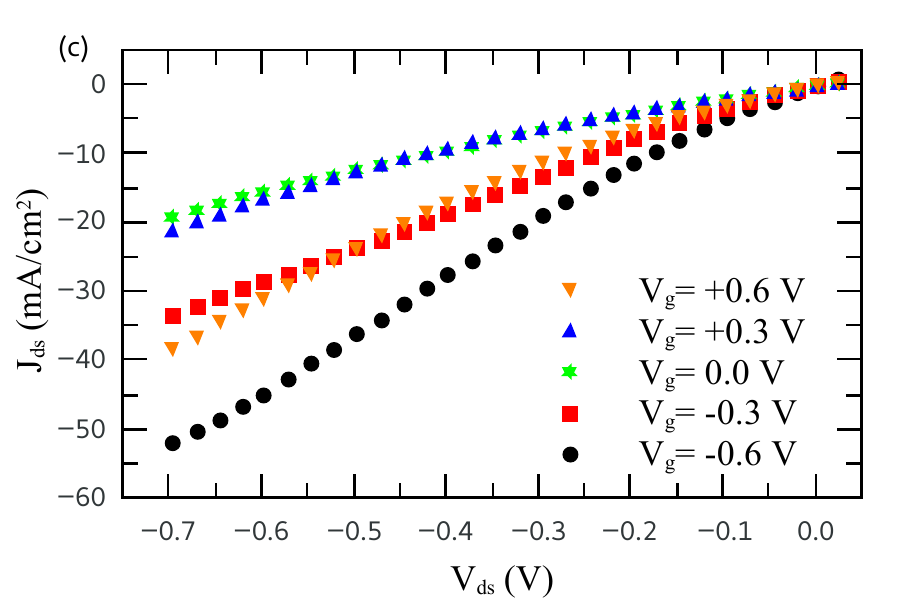}
\includegraphics[width=7.9cm,height=4.4cm]{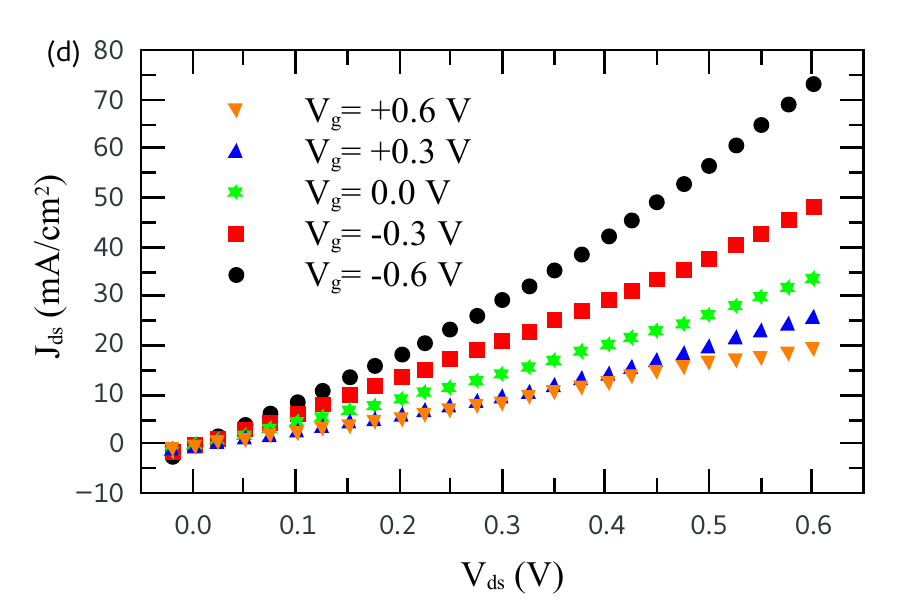}
    \caption{VET electrical characterization for gate voltage range B and ambipolar output current -- Transfer curve for ambipolar gate biasing depicting the output current density ($J_{ds}$--red circles) and leakage current density ($J_{g}$--black squares) for (a) $V_{ds}=-0.4\,\rm{V}$ and (b) $V_{ds}=+0.4\,\rm{V}$. Characteristic curve for (c) negative and (d) positive $V_{ds}$ range with ambipolar gate bias}
    \label{output_H_voltage}
\end{figure}

 The VET ana\-lysis for gate voltage range B $(-0.6\,\rm{V} < V_{g,B} < +0.6\,\rm{V})$ is depicted in Figure \ref{output_H_voltage}. Its charge carriers regimes are different from that observed for gate voltage range A. The modulation occurs for ambipolar $V_{ds}$ as not observed for the lower gate voltage range. Once this higher gate vol\-tage (Range B) is applied, no more the unipolar output current beha\-vior observed in the voltage range A is obtained. 
\par The transfer curves for gate voltage range B are depicted in Figure \ref{output_H_voltage}: (a) for $V_{ds}=-0.4\,\rm{V}$ bias and (b) for $V_{ds}=+0.4\,\rm{V}$ bias. When the applied gate voltage is $V_g>0.3\,\rm{V}$, cations diffuse inside the P3HT film changing its electrical properties and increasing its conductivity with the ion doping process \cite{ion_doping}. So, due to this ionic current through the IE pores/P3HT interface, this VET mode operation was named here as a Vertical Organic Electrochemical Transistors (VOECT). The presence of the ionic current makes this transistor stable no longer than five cycles of measurements initialising its degradation after that. However, this new device is stable when explored only in the gate voltage range A as an Electrolyte-Gated VOFET. More than twenty cycles of measurements were performed without any change when it operates as Electrolyte-Gated VOFET only. 
\par The characteristic curves for gate voltage (Range B) are depicted in Figure \ref{output_H_voltage}: (c) for negative $V_{ds}$ and (d) for positive $V_{ds}$ ranges. For negative $V_{ds}$ range, the output current profile is similar to observed in Range A. But, when $V_g>0.3\,\rm{V}$ is applied and cations diffuse through the IE pores, a magnification of the output current intensity is observed. Therefore, for negative $V_{ds}$ and positive $V_g$ range B is set, this VET works as a VOECT with modulation occurring due to the ionic current. However, when negative $V_{ds}$ and negative $V_g$ range B is set, this VET works as an Electrolyte-Gated VOFET due to its modulation being the result from induced charges in the channel. 
\par In Figure \ref{output_H_voltage}(d) is depicted the characteristic curve for positive $V_{ds}$. The off-state is formed when positive gate voltage induces electrons in the channel/IE pores anode. As a majority holes charge carriers device, this electrical configuration set closes these IE pores/channel interface transport. The on-state is formed when negative gate voltage is applied and open the IE pores/channel in the anode.
\par The two similar VETs that have already been reported in the literature \cite{ VET_2_Lenz, VET_2018} present higher On-Off ratio than the present VET but for a voltage range up to 3 \cite{VET_2_Lenz} to 20 \cite{VET_2018} times higher that in the present work does not becoming a comparable parameter. Here in this work, the main focus to be stressed is regarding to its different charge carriers transport regimes possible to be explored in VETs, never reported in this architecture before. The VET presented in this work has two distinctive mode operation as Electrolyte-Gated VOFET or VOECT depending on the gate voltage bias and range. For both mode of operation it is  possible to obtain a high On-state current density with intensity of $\sim 10^1 \,\rm{mA/cm^2}$.
%
\par Since that is a new device class based on VOFETs and EGOFETs architectures, it was compared our results with these two transistors efficiency and modes of operations. For a comparison with VOFETs: the great advantage of the present device is that the applied voltage range is much smaller since some of the best results for VOFETs is in the order of $V_{ds}=|1.5|\,\rm{V}$ and the current density is comparable to some VOFETs operating with voltage range of $V_{ds}\geq |2.0|\,\rm{V}$. On the other hand, when compared to EGOFETs architecture: the applied voltage range is similar in both architectures but the current density is much higher in our device. In this point of view we have in just one device architecture good features like: a active area of $\sim 9\,\rm{mm^2}$ with high current density been operating by a very low voltage. With applied voltage up to $0.5\,\rm{V}$, the presented current density is enough to drive a current able to generate luminescence as required in optoelectronic devices that normally requires higher voltage range \cite{Low_Voltage_Low-Power_Organic_Light_Emitting_Transistors,Semi_transparent_vertical_organic_light_emitting_transistors}. This situation has already been explored as a Vertical Electrolyte Gated Polymer Light-Emitting Transistor \cite{Electrolyte_Gated_Polymer_Light_Emitting_Transistor_2016} since the structure is very promising for applications in for low-power optoelectronic circuit.
Other possible applications are related to bio-sensor where the larger active area can ge\-nerate higher sensitivity for sensors or iontronic delivery devices \cite{bioelectronics_Malliaras} with larger active area when compared to EGOFETs or OECTs architectures \cite{Iontronic_Delivery_Device}.
\par In conclusion, here it is successfully reported a vertical electrolyte transistor (VET) based on a combination of two well known devices that are the: VOFET, exploring its stacked layers architecture and; the EGOFET, exploring its high capacitance of an ion gel dielectric structure. This combination results in a transistor able to drive high current density like in VOFETs with a very low voltage range operation like in EGOFETs. Depending on the gate voltage range applied, the transconductance in this transistor occurs: (i) due to charge carriers induced in the channel working as an  Electrolyte-Gated VOFET or; (ii) due to ionic current creating a VOECT. This VET architecture brings the possibility to work with transistors with very low voltage range simultaneously with a high current density.
\par We believe that further advance in this new architecture can improve research areas such as iontronic delivery devices and (bio-)sensors technology based on electrolyte solution. The possibility to develop VETs with larger active area can improve important characteristics of devices such as, the sensitivity of sensors, increase the transfer area for delivery drug applications. Finally, this VET shows great potential for implementation in low-power optoelectronic circuits.
\begin{table}[h]
\scriptsize
\begin{tabular}{|c|c|}
\hline
\multicolumn{2}{|c|}{\textbf{List of abbreviation}} \\ \hline
           EGOFET    &   Electrolyte Gated Organic Field Effect Transistor             \\ \hline
       OECT
        & Organic Electrochemical Transistor               \\ \hline
      VOFET
         &  Vertical Organic Field Effect Transistor              \\ \hline
         VET
      &   Vertical Electrolyte Transistor             \\ \hline
          Electrolyte-Gated VOFET
     & 
     \begin{tabular}[c]{@{}c@{}}Electrolyte-Gated\\ Vertical Organic Field Effect Transistor\end{tabular}
     \\ \hline
            VOECT
        &  Vertical Organic Electrochemical Transistor              \\ \hline
\end{tabular}
\end{table}
\section*{Acknowledgements}
\par The author would like to thank the Supramolecular Systems Research Group at Humboldt-Universit\"at zu Berlin for the hospitality during her sabbatical year providing the possibility to develop this research. I also would like to thank Felix Hermerschmidt, Simon Dalgleish and Emil J.W. List-Kratochvil at Humboldt-Universit\"at zu Berlin for fruitful comments and discussion and also for laboratory stuffs' financial support. 
%







\bibliographystyle{apsrev4-1}


%
\bibliography{cas-refs}


\end{document}